\documentclass[draft]{agujournal2019}
\usepackage{url} 
\usepackage{lineno}
\usepackage[inline]{trackchanges} 
\usepackage{soul}
\usepackage{amsmath}

\draftfalse

\journalname{JGR: Machine Learning and Computation}

\begin{document}

\title{Uncertainty Quantification in Reduced-Order Gas-Phase Atmospheric Chemistry Modeling using Ensemble SINDy}

\authors{Lin Guo\affil{1}, Xiaokai Yang\affil{1},  Zhonghua Zheng\affil{2}, Nicole Riemer\affil{3}, Christopher W. Tessum\affil{1}}

\affiliation{1}{Department of Civil and Environmental Engineering, University of Illinois Urbana-Champaign, Urbana, IL, USA}
\affiliation{2}{Department of Earth and Environmental Sciences, The University of Manchester, Manchester, United Kingdom}
\affiliation{3}{Department of Climate, Meteorology, and Atmospheric Sciences, University of Illinois Urbana-Champaign, Urbana, IL, USA}

\correspondingauthor{Christopher W. Tessum}{ctessum@illinois.edu}

\begin{keypoints}
\item Using a simple tropospheric ozone chemistry model, we quantify uncertainty caused by simplifying the model with machine learning.
\item Compared to the deterministic simplification method, the probabilistic model also reduces error.
\item Full-scale use could improve uncertainty quantification in atmospheric modeling, improving atmospheric insight and air quality control.
\end{keypoints}

%
%

%
%


\begin{abstract}

Uncertainty quantification during atmospheric chemistry modeling is computationally expensive as it typically requires a large number of simulations using complex models.
As large-scale modeling is typically performed with simplified chemical mechanisms for computational tractability, we describe a probabilistic surrogate modeling method using principal components analysis (PCA) and Ensemble Sparse Identification of Nonlinear Dynamics (E-SINDy) to both automatically simplify a gas-phase chemistry mechanism and to quantify the uncertainty introduced when doing so.
We demonstrate the application of this method on a small photochemical box model for ozone formation.
With 100 ensemble members, the calibration $R$-squared value is 0.96 among the three latent species on average and 0.98 for ozone, demonstrating that predicted model uncertainty aligns well with actual model error.
In addition to uncertainty quantification, this probabilistic method also improves accuracy as compared to an equivalent deterministic version, by $\sim$60\% for the ensemble prediction mean or $\sim$50\% for deterministic prediction by the best-performing single ensemble member. 
Overall, the ozone testing root mean square error (RMSE) is 15.1\% of its root mean square (RMS) concentration.
Although our probabilistic ensemble simulation ends up being slower than the reference model it emulates, we expect that use of a more complex reference model in future work will result in additional opportunities for acceleration.
Versions of this approach applied to full-scale chemical mechanisms may result in improved uncertainty quantification in models of atmospheric composition, leading to enhanced atmospheric understanding and improved support for air quality control and regulation.

\end{abstract}

\section*{Plain Language Summary}
To quantify the uncertainty that originates from simplifying complex atmospheric gas phase chemical mechanisms, we apply a probabilistic machine-learning framework (E-SINDy) to build a surrogate model that consists of multiple models trained with different subsets of data and species. As demonstrated on a simple photochemical mechanism, this method can effectively and reliably quantify the uncertainty in its predictions and shows promise toward scaling to more complicated atmospheric models. Compared to an equivalent deterministic approach, E-SINDy is not only more robust but also more accurate when predicting the levels of various substances in the atmosphere under different environmental conditions. With a full-scale reference mechanism, this method could greatly improve uncertainty quantification in atmospheric modeling, enhancing scientific ability to understand atmospheric changes and supporting air quality control.

\section{Introduction}

Mathematical modeling of atmospheric chemistry integrates diverse scientific disciplines---meteorology, radiative transfer, physical chemistry and biogeochemistry---allowing for a comprehensive understanding and quantification of the factors controlling atmospheric concentrations of chemicals and their interrelated processes, thereby supporting both scientific advancement and informed policy-making in air quality management \cite{brasseur2017AQMtextbook}. 
However, this modeling is computationally challenging, owing to the large number of chemical species in the atmosphere and the numerical stiffness that results from the disparate time scales at which the dynamics of different species occur \cite{brasseur2017AQMtextbook, shen2022machine}.
Uncertainty in atmospheric chemistry simulations can include structural uncertainty from the discrepancy between the model architecture and actual causal relationships, uncertain input data, uncertain physical or chemical constants (parameter uncertainty), and numerical uncertainty due to discretization and rounding.
Although each of these uncertainties can be substantial, most atmospheric chemical transport model (CTM) applications do not formally quantify uncertainty at all \cite{aleksankina2019advanced}. 
However, uncertainty quantification is important for model validation and reproducibility, providing critical context for model results \cite{volodina2021importance}.
Specifically for atmospheric chemistry modeling involving gas-phase chemistry, uncertainty quantification provides an assessment of confidence and provides distributions of the predicted chemical concentrations, enhancing the interpretability of the results in support of policy analysis and management \cite{aleksankina2019advanced, kashinath2021physics}.

Traditionally, the Monte Carlo (MC) method has been most widely applied for parametric uncertainty analysis of atmospheric modeling \cite{derwent1988application, chen1997chemical, hanna1998monte, hanna2001uncertainties}, along with its combination with a stratified sampling strategy such as Latin hypercube sampling to reduce computational effort \cite{derwent1988application, bergin1999formal} and with its extension by introducing Bayesian techniques to reduce the affect of subjective priors (``Bayesian MC''; \cite{bergin2000application, beekmann2003monte}).
However, the computational cost of MC-based approaches are often prohibitively high because they require a large number of simulations corresponding to a large number of samples in the parameter space \cite{aleksankina2019advanced, huang2019feasible}. Thus, the use of MC is computationally impractical in many use cases, especially with complex models \cite{castelletti2012general}.
Therefore, previous applications of the MC method have applied simple air quality models, narrow geographic regions, or limited parameter spaces to maintain manageable computational costs.
For example, \citeA{bergin1999formal} applied Latin hypercube sampling with MC on the uncertainty quantification of the trajectory version of the California/Carnegie Institute of Technology air quality model, which is more computationally tractable than three-dimensional models.
\citeA{beekmann2003monte} apply the Bayesian MC approach to address CHIMERE \cite{schmidt2001comparison} model input and parameter uncertainties in the Paris urban area with fixed boundary and initial conditions.

To a certain extent, this computational intensity of MC can be alleviated by surrogate modeling—the use of machine learning to create simplified models that emulate the behavior of more complex models.
For example, \citeA{aleksankina2019advanced} adopt Gaussian process emulator as a non-parametric surrogate model on the WRF-EMEP4UK model \cite{vieno2010modelling, vieno2014role,vieno2016sensitivities}.
It is relatively skilled with a small dataset \cite{conibear2022sensitivity} but its computational intensity increases cubically with the dataset size \cite{barber2012bayesian}.

The literature cited above has studied the effect of uncertainty in model inputs or model parameters on model outputs, thus making the implicit assumption that the structure of the model is correct \cite{smith2013uncertainty}.
However, to achieve computational tractability, even the most advanced 3D chemical transport models such as GEOS-Chem \cite{bey2001geoschem}, CMAQ \cite{byun2006review}, or CAMx \cite{environ2014user} typically include chemical reaction networks that are greatly simplified compared to ``full-complexity'' models such as MCM \cite{jenkin1997mcm1,saunders2003mcm2} and GECKO \cite{aumont2005modelling}. 
This model reduction introduces structural uncertainty: even if scientists were able to create an extremely high-fidelity model of atmospheric chemistry (we are not quite there yet), any reduced-order version of this model that is simplified enough to routinely run in a 3D simulation would be structurally different than the original, and this structural uncertainty has never been rigorously quantified. 
Here, we will describe a method for quantifying the uncertainty introduced when simplifying an atmospheric gas-phase mechanism for the purpose of increasing its computational speed, both in terms of the structure of the equations of the model and in terms the parameters or coefficients applied to each equation term.
(For the analysis here, we assume that there is no uncertainty in model input data including initial concentrations, emissions, temperature, pressure, and solar radiation.)

Traditional gas-phase atmospheric chemistry model reduction has relied on expert intuition to manually combine multiple explicit chemical species into representative groups \cite{stockwell2011review, carter2010development}.
Recently, surrogate modeling has shown potential as a more automated method for atmospheric chemistry model order reduction~\cite{keller2017machine,keller2019application,kelp2020rnn, sturm2022conservation,schreck2022,kelp2022online,huang2022neural}.
These surrogate models have typically struggled to maintain numerical stability over long simulations, but in recent work we have demonstrated the development of an accurate, numerically stable surrogate model of a small-scale gas-phase atmospheric chemistry system which achieves a substantial speed-up over its reference model \cite{xk2024surrogate}.
Below, we will demonstrate the quantification of the uncertainty introduced during the model reduction process described in our previous work \cite{xk2024surrogate}.

\section{Data and Methods}
\subsection{Reference Model and Data}
To investigate the performance of our surrogate modeling technique, we build a reference model with which we generate data for training our surrogate models. Our reference model setup and data generation are described in detail by \citeA{xk2024surrogate}. In brief, it is a simple photochemical box model focused on simulating the dynamics of gas-phase tropospheric atmospheric chemistry as described by \citeA{sturm2020mb}.
It comprises 10 reactions of 11 species (Table~S1).
To improve the realism of our simulations, we add external forcing factors such as solar radiation, emissions, and deposition so that the system tends to follow the ``diurnal ozone cycle'' characteristic of tropospheric chemistry \cite{jacob1999introduction}.  
We use Sobol sampling to randomly vary the time-of-day at the beginning of the simulation as well as temperature, pressure, emissions patterns, and peak radiation intensity which is represented by the cosine of solar zenith angle, thereby generating 3000 three-day concentration trajectories for training, 375 ten-day trajectories for validation, and 375 ten-day trajectories for testing.
The emission flux and radiation intensity are set to a diurnal pattern, and the temperature and pressure are set constant along each trajectory.
We discard the first day of simulation to remove the effect of the initial conditions.
Simulated species concentrations range in magnitude from $10^{-10}$ to $10^1$~ppm. 
Following \citeA{xk2024surrogate}, we preprocess the simulation data by subtracting the mean concentration for each species and performing principal component analysis (PCA) \cite{wold1987principal, brunton2017chaos, conti2023} on the result and we retain the three principal components for dimension reduction, which we consider as sufficient as they represent over 85\% of the total variance in the original system \cite{xk2024surrogate}. 
We refer to these principal components as ``latent species''.
This PCA dimensionality reduction allows our surrogate model to represent chemical dynamics using a small number of state variables, thus reducing computational cost \cite{champion2019data, bakarji2022discovering}; related theory and implications are described in detail by \citeA{xk2024surrogate}.

\subsection{Surrogate Modeling with deterministic SINDy}
Our approach is based on the Sparse Identification of Nonlinear Dynamics (SINDy) \cite{sindy} framework, which has shown promise in emulating the dynamics of differential equation systems in a wide variety of fields \cite{lai2019sparse,hoffmann2019reactive,wang2021inference,jiang2021modeling,pasquato2022sparse}.
Our deterministic SINDy method is described in detail by \citeA{xk2024surrogate}; we describe it briefly here.
With a provided library of possible equation terms, SINDy can identify a differential equation system that can explain the dynamics in a data set by balancing the number of equation terms in the surrogate model against the accuracy of predictions. 
The optimization problem solved by deterministic SINDy is shown in Equation \ref{eq:sindy}:
\begin{linenomath*}
\begin{equation}
 \mathop{\mathrm{argmin}}_\Xi \| \dot{C} - \Theta\left(C,p\right)\Xi \|_2 + \lambda_{\mathrm{threshold}} \|\Xi\|_0 + \lambda_{\mathrm{ridge}} \|\Xi\|_2,  \label{eq:sindy}
\end{equation}
\end{linenomath*}
where $\dot{C} \in \mathbf{R}^{t \times d}$ is the derivative of each latent species $C$ in the training dataset with respect to time, $\Theta(C,p) \in \mathbf{R}^{t \times k}$ is the candidate equation term library constructed from $C$ and $p$, and $\Xi \in \mathbf{R}^{k \times d}$ holds the coefficients corresponding to the equation terms in $\Theta(C,p)$. Equation term coefficients with values less than $\lambda_{\mathrm{threshold}}$ are set to zero.
$\lambda_{\mathrm{ridge}}$ is the coefficient for ridge regression and defaults to a value of 0.05.
The total number of timesteps in the training dataset is denoted by $t$. 
The number of equation terms in the constructed library is denoted by $k$, and $d$ is the number of latent species after PCA.
The system parameters including pressure ($P$), temperature ($T$), emissions ($E$), deposition flux ($D$), and radiation intensity ($h\nu$) are represented as $p \in \left\{ P, T, E, D, h\nu \right\}$.

When using deterministic SINDy, the equation term library $\Theta\left(C,p\right)$ needs to be selected.
The selection should balance the need for including all potential terms which may be needed to explain the dynamics with the possibility of including so many terms that the optimization problem in Equation \ref{eq:sindy} becomes difficult to solve.
Typically this is done by considering the type of system to be modeled and including the terms that could be expected to appear in that system.
Therefore, here we select equation library terms similar to those found in chemical reaction kinetics, namely polynomial terms by themselves and also multiplied with terms in $p$, as shown in Equation \ref{eq:sindy_library}:
  \begin{linenomath*}
  \begin{equation}
\Theta(C,p) = \left\{ C_1, C_1C_2, C_1^2, ..., C_d^n; \left(C_1, C_1C_2, C_1^2, ..., C_d^n \right) \cdot  \left[ PT, h\nu \right]; PT, E, D, h\nu \right\}, \label{eq:sindy_library}
 \end{equation}
  \end{linenomath*}
where $C_i$ is the concentration of the $i^\textrm{th}$ of $d$ latent species and $n$ is the maximum power for the polynomial basis. 
$n$ is chosen as 4 by hyperparameter tuning
and the number of equation terms $k=113$.
After constructing the matrix $\Theta$ from the terms in Equation~\ref{eq:sindy_library}, we solve Equation \ref{eq:sindy} using the Sequentially Thresholded Ridge Regression (STRidge) algorithm \cite{rudy2017data, fasel2022ensemble} which performs repeated regressions, setting values in $\Xi$ below a user-specified threshold $\lambda_{\mathrm{threshold}}$ to zero after each regression and repeating until convergence.
Like \citeA{xk2024surrogate}, we focus on predicting ozone concentration, applying an ozone weight coefficient $\beta$ to increase the prominence of ozone in the resulting surrogate model.

\begin{figure}[h]
\centering
 \includegraphics[width=\textwidth]{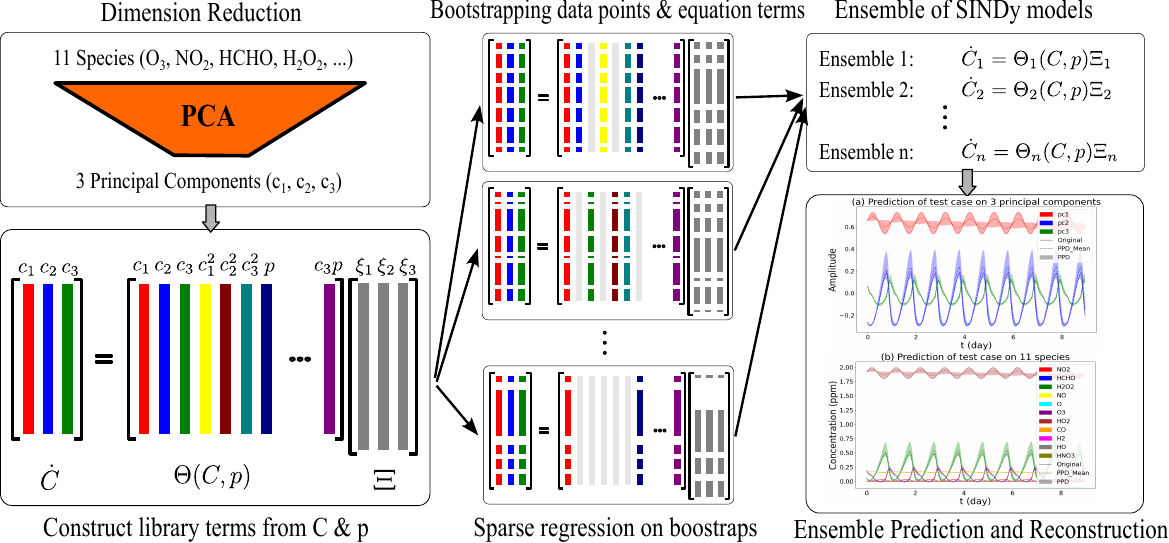}
 \caption{Overall schematic of creating an Ensemble SINDy-based surrogate for a gas-phase atmospheric chemical mechanism.}
 \label{fig:esindyworkflow}
 
 \end{figure}

\subsection{Probabilistic Surrogate Modeling with Ensemble SINDy} \label{sec:esindy}
Here, we build on the work of \citeA{xk2024surrogate} by including uncertainty quantification.
For this, we employ Ensemble SINDy (E-SINDy; \cite{fasel2022ensemble}), which generates ensemble member models for making predictions and quantifying uncertainty \cite{gao2023convergence}. 
This approach leverages bootstrapping techniques applied to the candidate equation terms of the library and the observations in the training data, as illustrated in Figure~\ref{fig:esindyworkflow}. 
This method not only inherits the rapid training and inference speed and interpretable models of deterministic SINDy for each ensemble member, but also provides a probabilistic distribution for each coefficient along with confidence interval (CI) for predictions.
To obtain $n$ posterior samples, its computational complexity is lower with a lower bound of $O(n^2)$ than the costly Bayesian Markov chain Monte Carlo uncertainty quantification method with an upper bound of $\Omega(n^3)$ \cite{gao2023convergence}.
In this study, we implement the E-SINDy framework as described by \citeA{fasel2022ensemble}.
Specifically, we:
\begin{enumerate}
    \item Subset the data and features by repeatedly sampling with replacement from the fraction of training data $\dot{C}$ and the candidate equation terms of the constructed library $\Theta(C, p)$; 
    \item Perform sparse regression on each subset of the sampled data and features, resulting in various ensemble members coefficient sets, denoted as $\Xi_b$; and
    \item Optionally select a subset of the ensemble members that perform best on the training data for making predictions and quantifying uncertainty.
\end{enumerate}

We treat the number of data points and equation library terms sampled for bootstrapping as hyperparameters, which we optimize as described in Section \ref{sec:implementation}. 
We also try other sampling strategies, including experimenting with data bagging and library bagging separately. 
However, these experiments do not result in improved accuracy compared to the results shown here.
During the model training for each ensemble member, we find some members lack provisions for ensuring numerical stability for different training samples. 
Inspired by \citeA{uqsindy}, we explore adding a buffer term---a higher-order polynomial term with a small, negative weight coefficient $\epsilon$---into each equation to encourage numerical stability.
If the library of candidate functions includes polynomial terms up to order $n$, we add a term $-\epsilon x^{n+1}_i$ if $n$ is even, or $-\epsilon x^{n+2}_i$ if $n$ is odd.
Additionally, we explore the inclusion probability of each coefficient, which is the fraction of non-zero coefficients across the optionally selected ensemble members, and the effect of inclusion probability threshold $\lambda_{\mathrm{ip}}$, as discussed in Section \ref{sec:IP}.

\subsection{Implementation} \label{sec:implementation}
We use the E-SINDy implementation in the ``pysindy.py'' software library \cite{kaptanoglu2021pysindy}. 
During the inference phase, the solver LSODA \cite{hindmarsh2005lsoda} is used within the ``numbalsoda.py'' software package. 
Instead of using the equation term library described by \citeA{xk2024surrogate}, we start with similarly structured libraries of fourth order polynomials instead of third order polynomials, as shown in Equation \ref{eq:sindy_library}: [fourth order polynomial of C, Emissions, Deposition, fourth order polynomial of C $\times$ radiation intensity, fourth order polynomial of C $\times$ pressure $\times$ Temperature], as we find that this selection of candidate terms works better for our E-SINDy case.
For the hyperparameter tuning, we use $\left [30\%, 60\%, 90\% \right]$ for the percentage of data sampled and $\left [30, 60, 90\right]$ for the number of candidate equation terms dropped, where there are a total of 113 candidate equation terms under the current library setting. 
We select 30\% of the data points and 30 equation terms to drop for each ensemble, and tune the buffer term coefficients $\epsilon \in \left\{0, 10^{-9}, 10^{-8}, 10^{-7}, 10^{-6}, 10^{-5}, 10^{-4}, 10^{-3}, 10^{-2}, 10^{-1}\right\}$ to optimize trade-offs between stability and accuracy during time-series inference on the validation dataset. 
We find the buffer term cannot increase the stability effectively without sacrificing accuracy, so we therefore choose the value of zero.
For the inclusion probability threshold we try $\lambda_{\mathrm{ip}} \in \left\{0, 0.5, 0.7\right\}$ and find zero is optimal, meaning that our final model does not exclude any equation terms based on their inclusion probability.
For other hyperparameters, empirical analysis indicates that a value of 1.48 for the ozone weight coefficient $\beta$ and a value of $3.3\times10^{-5}$ for the threshold parameter $\lambda_{\mathrm{threshold}}$ are effective.

With this optimal set of hyperparameters, we obtain 1000 ensemble members and select the top 10\% ensemble members with the lowest training errors on O\textsubscript{3} for ensemble forecasting using the posterior predictive distribution (PPD).
Although our use of only some of the original ensemble members to calculate the PPD is not entirely consistent with the empirical distribution function and the plug-in principle used in bootstrapping analysis \cite{efron1994introduction, hall2013bootstrap}, it appears that some samples of the candidate equation term library do not contain all of the terms necessary to create a reasonable model of the dynamics, and therefore removing the worst performing members from the ensemble results in reduced ensemble error without adversely affecting uncertainty calibration, as shown in Section \ref{sec:results}.
We also choose a ``best ensemble member'' that demonstrates the lowest training error for O\textsubscript{3} and exhibits 100\% stability across all training cases for single-model deterministic prediction.  

\section{Results} \label{sec:results}
To examine model performance, we initially focus on a single case to illustrate how E-SINDy makes probabilistic predictions using the PPD (Section \ref{model_performance}). 
We then broaden our discussion to explore the qualitative variability in performance across different cases for predictions of ozone.
We also highlight how E-SINDy improves efficiency in ensuring model stability and improves accuracy as compared to our previous results \cite{xk2024surrogate}.
Section \ref{calibration} further explores the reliability of the E-SINDy uncertainty quantification, while Section \ref{computational_speed} examines the computational efficiency. 
Section \ref{hyperparameter} explores the hyperparameter tuning process during bootstrapping. 
Section \ref{buffer_term} examines the impact of the buffer term coefficient $\epsilon$ on model stability and performance. 
Lastly, Section \ref{sec:IP} investigates the influence of the inclusion probability threshold $\lambda_{\mathrm{ip}}$ and aggregated model inference, which are part of the original E-SINDy method \cite{fasel2022ensemble}.

\subsection{Model Performance} \label{model_performance}
\begin{figure}[h]
\centering
 \centerline{\includegraphics[width=\textwidth]{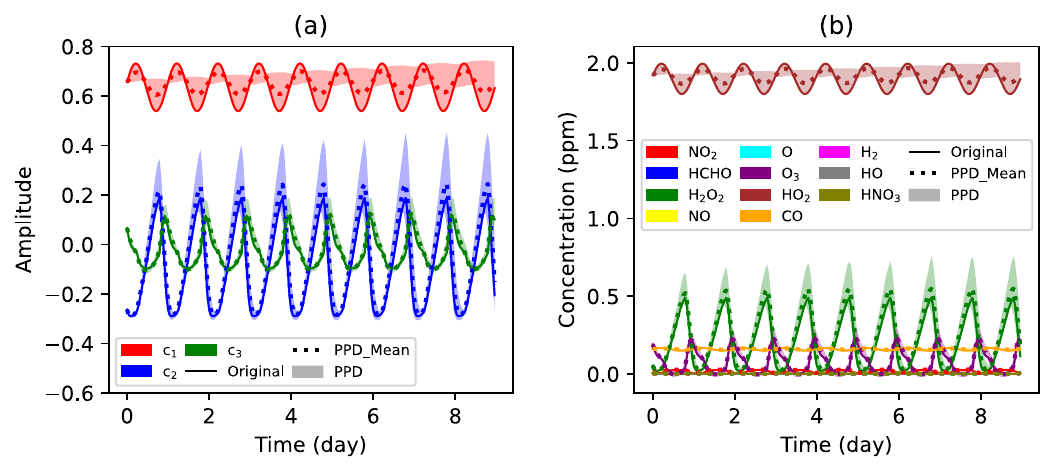}}
 \caption{(a) Prediction of three latent species concentrations for an example nine-day testing case. (b) Prediction of concentrations of 11 species for nine-day testing case (The ``PPD'' or shaded area denotes the 95\% CI of our surrogate model ensemble, which comes from the predictions of the 100 selected ensemble members with the lowest training errors on O\textsubscript{3}. ``PPD\textunderscore Mean'' denotes the mean prediction from the PPD.)}
 \label{fig:testcase}
 \end{figure}

To quantify uncertainty, we select the 100 ensemble members with the lowest training error on O\textsubscript{3} to perform inference on a single case. 
Subsequently, we extract the posterior mean prediction and the 95\% CI from the ensemble PPD.
Figure~\ref{fig:testcase} demonstrates an ensemble prediction for a randomly chosen case from the test dataset using three latent species (Figure~\ref{fig:testcase}a), and also the result after decompressing the simulation result back to the 11 species of the original reference model (Figure~\ref{fig:testcase}b). 
Qualitatively, the mean of the posterior distribution closely tracks the reference model prediction, although some of the ensemble members do not capture the diurnal pattern in the first latent species.
The ``true'' reference model value falls within the 95\% CI of our surrogate model ensemble the vast majority of the time, both for the latent species and the reconstructed original species concentrations.

 \begin{figure}[h]
\centering
 \centerline{\includegraphics[width=\textwidth]{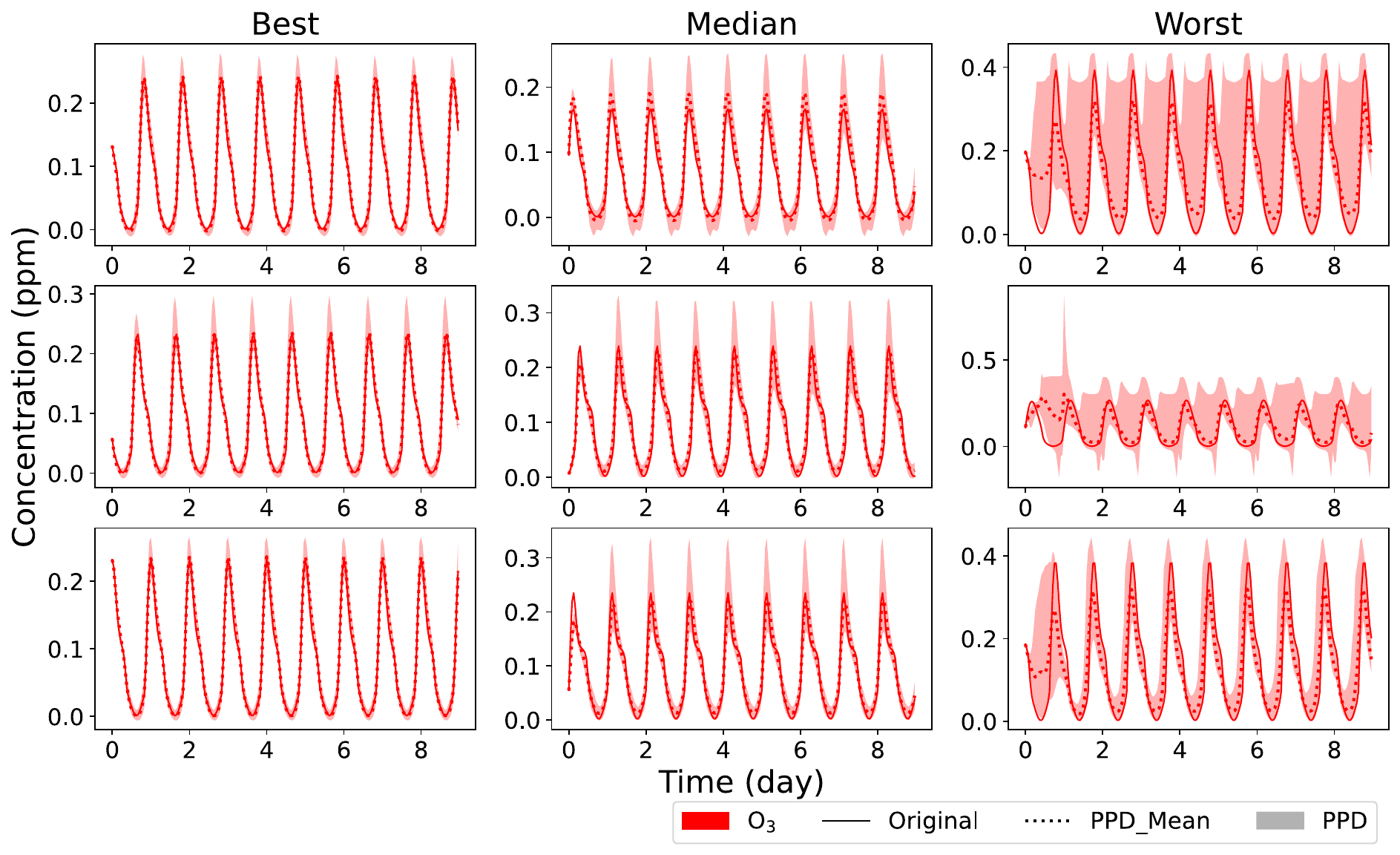}}
 \caption{The three best, median and worst predictions of O\textsubscript{3} concentrations among the 375 nine-day testing cases. The ``PPD'' shaded area denotes the 95\% CI, which comes from the predictions of the 100 selected ensemble members with lowest training errors on O\textsubscript{3}. ``PPD\textunderscore Mean'' denotes the mean of the PPD.}
 \label{fig:bestmedianworst}
 \end{figure}

 \begin{figure}[h]
\centering
\centerline{ \includegraphics[width=\textwidth]{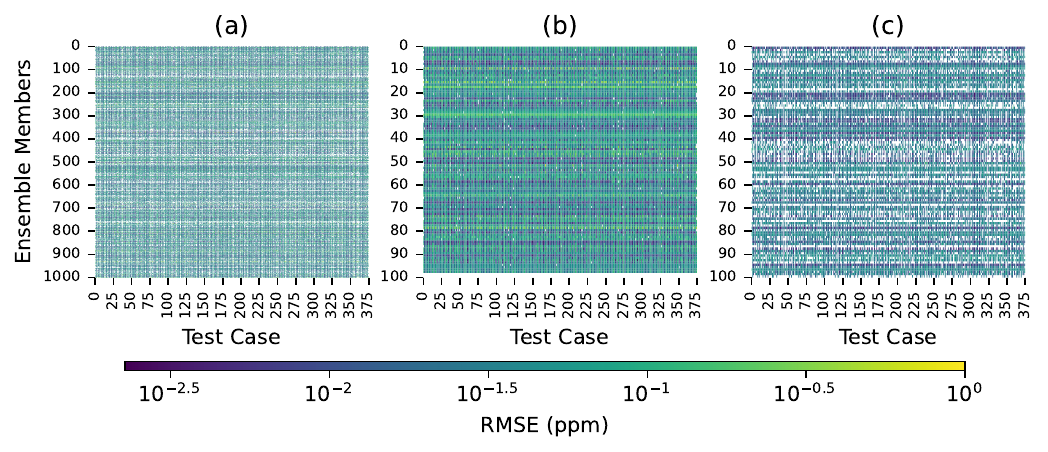}}
 \caption{(a) Testing RMSE (ppm) by 1000 ensemble members on 375 testing cases (white denotes unstable solutions) (b) Testing RMSE (ppm) by ensemble members able to stably solve 95\% of testing cases (c) Testing RMSE (ppm)  by selected 100 ensemble members with lowest training errors on O\textsubscript{3}.}
 \label{fig:testingRMSEs}
 \end{figure}
 \begin{figure}[h]
\centering
 \centerline{\includegraphics[width=\textwidth]{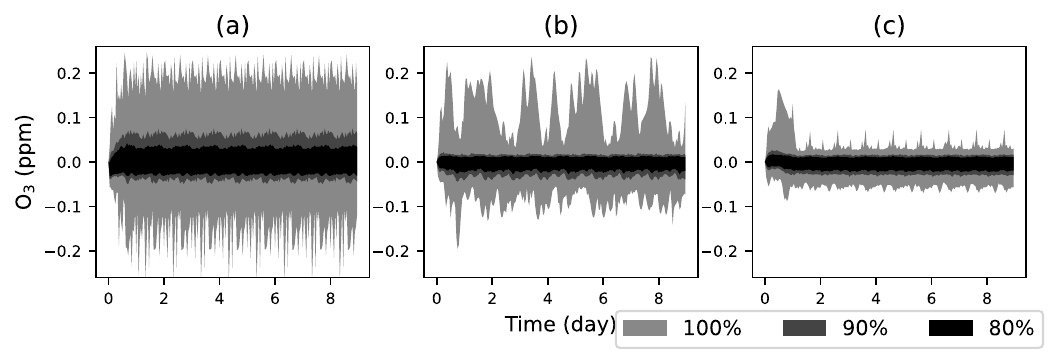}}
 \caption{Error percentiles for predictions of O\textsubscript{3} concentration for 375 testing cases by (a) deterministic SINDy, (equivalent to results from \citeA{xk2024surrogate}); (b) the best E-SINDy ensemble member; and (c) the PPD mean of the 100 selected ensemble members. ``Best'' model and ``selected'' models are selected based on training data, not testing data.}
 \label{fig:epercentile}
 \end{figure}

Having demonstrated the outcome of a single simulation, we can now examine the variation in performance across multiple simulations.
For uncertainty quantification across all testing cases, 89\% of the reference values for the three latent species fall within the 95\% CI.
Among the reconstructed 11 species, 60\% of the reference values are encompassed within this interval.
Specifically for O\textsubscript{3}, the proportion of reference values included within the CI reaches 91\%.
Figure~\ref{fig:bestmedianworst} shows this variability qualitatively. 
For O\textsubscript{3}, the best 3 cases have the lowest testing RMSE of 0.0024~ppm, 0.0026~ppm, and 0.0026~ppm, which are 2.7\%, 2.9\% and 2.9\% of RMS of O\textsubscript{3} concentration (0.090~ppm). 
For median cases, the RMSEs are 0.0075~ppm, which is 8.3\% of the RMS. 
For worst cases, the RMSEs are 0.0297~ppm, 0.0299~ppm and 0.0309~ppm, which are 32.9\%, 33.1\% and 34.2\% of the RMS.
The highest RMSE in these ``worst cases'' is 58.8\% lower than that in our previous work with deterministic SINDy \cite{xk2024surrogate}, demonstrating the potential for probabilistic surrogate models to increase model accuracy and robustness in addition to quantifying uncertainty.

The use of multiple ensemble members addresses a frequent challenge encountered in deterministic surrogate models \cite{ keller2017machine,kelp2018orders,keller2019application,kelp2020rnn,kaptanoglu2021promoting,schreck2022,kelp2022online,huang2022neural}, which is also apparent here for individual ensemble members: they may exhibit numerical instability or ``mean drift'' for some testing cases (although our previous work in \citeA{xk2024surrogate} does not, owing to a carefully construction candidate equation term library).
Figure~\ref{fig:testingRMSEs}(a) displays the RMSE on each testing case from each of our 1000 ensemble member models $\Xi_b$, where the overall testing RMSE is 0.0449~ppm and a 23.1\% improvement over deterministic SINDy \cite{xk2024surrogate}.
Most cases that were successfully simulated show a small error scale with a deep blue color. 
White parts denote an unstable solution, indicating that even when using a buffer term we may not guarantee a numerically stable solution or prevent ``mean drift'' in all cases. 
However, a benefit of ensemble predictions is that simulations that are not successfully completed by one ensemble member may be stably and accurately solved by other ensemble members.
Analyzing the RMSE of those ensemble members achieving 95\% stability in testing cases (Figure~\ref{fig:testingRMSEs}b), we observe consistent patterns in RMSE across both the cases and the ensemble members, as demonstrated by a uniform color scale that aligns parallel to either the horizontal or vertical axis. 
In scenarios where ensemble members are chosen based on training accuracy, the resulting ``select'' ensemble, illustrated in Figure~\ref{fig:testingRMSEs}(c), exhibits fewer instances of high error on the testing data (such as the yellow color observed in Figures \ref{fig:testingRMSEs}(a) and (b)). 
Both the RMSE and stability manifest as parallel strips along the axes, highlighting that some ensemble members outperform others by delivering more stable solutions or lower testing errors and certain cases remain inherently more challenging to simulate.

While our primary objective is to quantify uncertainty, an advantageous byproduct of our methodology is the improvement in accuracy, which can be further enhanced through the selection of the most accurate ensemble members. 
From the 100 selected ensemble members, chosen for their lowest training errors on O\textsubscript{3}, we observe a PPD testing RMSE of 0.0136~ppm for O\textsubscript{3}, marking a 59.9\% improvement compared to deterministic SINDy. 
For the original 11 species, the average testing RMSE is 0.0295~ppm, indicating a 49.4\% improvement over deterministic SINDy. 
The average rate at which testing cases were stably solved by each ensemble member—the stable solution rate—stands at 65\%. 
Selecting the ensemble member that is stable across all training scenarios and exhibits the lowest training error for O\textsubscript{3} (0.0117~ppm) results in an RMSE of 0.0179~ppm on the testing data for O\textsubscript{3}, a 47.2\% improvement over deterministic SINDy, with the RMSE for the 11 species remaining at 0.0296~ppm with improvement of 49.3\% over deterministic SINDy.
Figure~\ref{fig:epercentile} presents a comparison of the error intervals for O\textsubscript{3} predictions between deterministic SINDy \cite{xk2024surrogate}, the best E-SINDy ensemble member, and the selected 100 ensemble members from E-SINDy across 375 testing cases. In comparison to deterministic SINDy, both the best ensemble member with the lowest training error for O\textsubscript{3} and the selected 100 ensemble members reduce error at all percentiles shown. 
This trend of enhancement is consistently observed across all eleven species, as illustrated in Figures S1—S11.
The error plot for the selected ensemble members highlights a higher error during the initial stage (first day) across the 100\% interval, suggesting that less than 10\% of the simulations in the testing dataset exhibit a relatively high error at the beginning of the simulation.
Upon narrowing the interval from 100\% to 90\%, the error markedly decreases and stabilizes. 
Further reduction of the interval to 80\% does not significantly alter the error, which remains steady at approximately 13 parts per billion (ppb) and exhibits a diurnal pattern.

\subsection{Calibration} \label{calibration}
\begin{figure}[h]
\centering
 \centerline{\includegraphics[width=\textwidth]{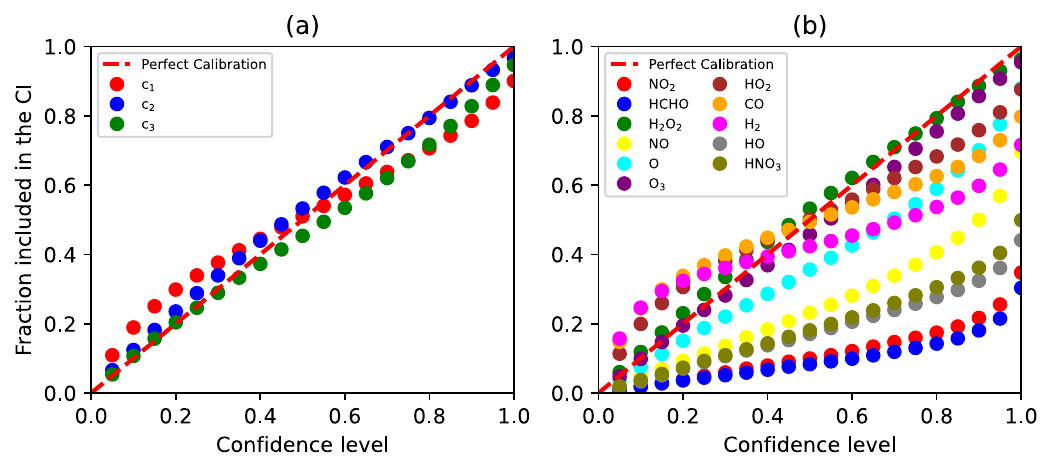}}
 \caption{Calibration curves showing the fraction of reference model predictions included in the PPD confidence interval as a function of confidence level for (a) three latent species and (b) 11 original species.  PPD results are from 100 selected ensemble members on 375 testing cases.}
 \label{fig:calibration}
 \end{figure}

A model designed to quantify uncertainty is only useful when its uncertainty estimates are accurately and reliably calibrated, meaning that the confidence interval of the PPD matches the probability of the true value falling within the PPD.
For example, a well-calibrated model with a 60\% CI would predict a CI that included the true value 60\% of the time.
Figure~\ref{fig:calibration} shows the calibration of our 100 selected ensemble members against the testing dataset.
Qualitatively, the latent species predictions are well calibrated, as are the predictions of O\textsubscript{3} (which is our focus here) and several other ``original'' species, but predictions of other original species are less-well calibrated.
We can also quantify calibration performance by calculating $R$-squared between the perfect 1:1 calibration line and each curve (``calibration $R$-squared'').
Doing so, we find that the curves for the three latent species have calibration $R$-squared values 0.927, 0.991 and 0.967 respectively.  
Also, the H\textsubscript{2}O\textsubscript{2}, O\textsubscript{3}, HO\textsubscript{2}, CO, O and H\textsubscript{2} are well calibrated with $R$-squared values of 0.992, 0.984, 0.899, 0.784, 0.748 and 0.600, respectively. 
For the other species, calibration $R$-squared values are lower, ranging from $-1.504$ to 0.052. 
However, we hypothesize that if we focused our surrogate modeling efforts on different pollutants the way we are currently focusing on O\textsubscript{3}, the calibration performance for those pollutants would increase (and the performance for O\textsubscript{3} would decrease).

 \begin{figure}[h]
\centering
 \centerline{\includegraphics[width=0.5\textwidth]{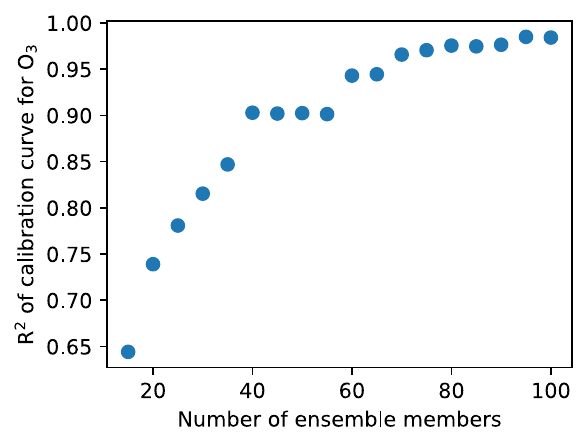}}
 \caption{$R$-squared value between O\textsubscript{3} calibration curve and perfect calibration vs. number of members included in the surrogate model ensemble.}
 \label{fig:r2calibration}
 \end{figure}

\subsection{Computational Speed} \label{computational_speed}
In order to quantify uncertainty, we perform simulations with multiple ensemble members. 
Leveraging the 10$\times$ speedup afforded by our surrogate model as compared to the reference model \cite{xk2024surrogate}, using 100 ensemble members results in an ensemble prediction that is 10$\times$ slower than a single prediction with the reference model. 
However, if we reduce the number of ensemble members to 30, the calibration $R$-squared value for O\textsubscript{3} is still above 0.8 (Figure~\ref{fig:r2calibration}), with the inference speed being only 3$\times$ slower than the baseline. 
Admittedly, a surrogate model that is slower than the reference model would not be beneficial for operational use. However, our goal here is to explore the characteristics of this approach using a small-scale reference model. 
When applying this approach to a larger reference model (of the type that would be used operationally) we would expect a larger speedup factor, because larger atmospheric chemical systems tend to include more-highly-correlated variables than the small system we use here, providing additional opportunities for compression.
From an accuracy standpoint, the deterministic prediction derived from the ensemble member with the lowest error and 100\% stability across all training scenarios maintains the same 10$\times$ speedup, while concurrently achieving a significant improvement in accuracy compared to our previous work \cite{xk2024surrogate}.
Similar to surrogate modeling with Random Forest \cite{keller2019application}, E-SINDy possesses a parallel structure, allowing each ensemble member to be trained or to perform inference concurrently. 
However, this aspect is not considered in this computational speed comparison because the reference model is also capable of being parallelized, for example across individual grid cells in a 3D model.

\begin{figure}[h]
\centering
 \centerline{\includegraphics[width=\textwidth]{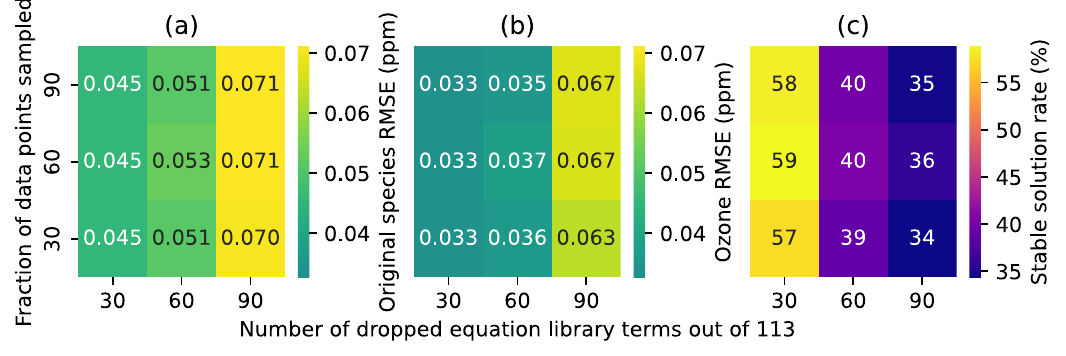}}
 \caption{The effect of fraction of data points and number of candidate equation terms sampled during bootstrapping on (a) average validation RMSE for original species, (b) validation RMSE for O\textsubscript{3}; and (c) stable solution rate.}
 \label{fig:bootstrap_tuning}
 \end{figure}
 
\subsection{Effect of Bootstrapping-Related Hyperparameter Choice}
\label{hyperparameter}
Figure~\ref{fig:bootstrap_tuning} shows the average original-species RMSE and the RMSE for O\textsubscript{3} (both averaged across all ensemble members and all cases in our validation dataset), as well as the stable solution rate, under different fractions of data points and candidate equation terms sampled from the data and library terms,  
Figure~\ref{fig:bootstrap_tuning}(a) and (c) show that the accuracies decrease with the increase of the number of dropped candidate equation terms.
When the number of dropped candidate equation terms is 90, the prediction becomes much less accurate due to the small size of the fitted model (which only consists of 23~terms). 
Figure~\ref{fig:bootstrap_tuning}(c) shows that the stable solution rate decreases with the increasing number of dropped candidate equation terms from 30 to 90. 
There is little difference in the accuracy and stable solution rate between different percentages of data points sampled, indicating that the size of the candidate term library during fitting is the dominant factor affecting accuracy and stable solution rate. 

  \begin{figure}[h]
\centering
 \centerline{\includegraphics[width=0.5\textwidth]{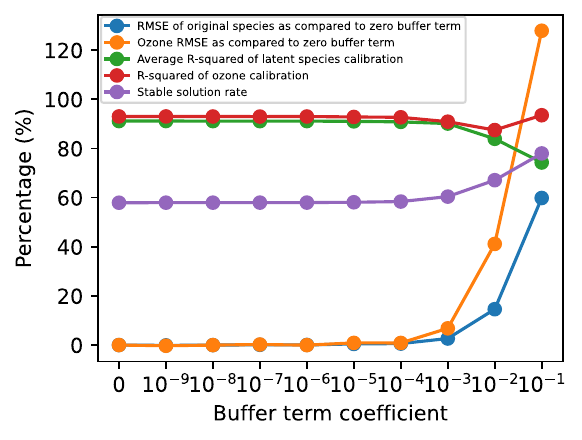}}
 \caption{The impact of choice in buffer term coefficient on model performance as measured by prediction error, stable solution rate, and uncertainty calibration.}
 \label{fig:buffer_term}
 \end{figure}

\subsection{Effect of Buffer Term} \label{buffer_term}
After choosing an optimal number of dropped candidate equation terms and percentage of sampled data points, we explore the effect of the buffer term coefficient $\epsilon$ for values ranging from zero to $10^{-1}$ on the performance of our 100 selected ensemble members on the validation cases. 
Specifically, we measure the percent change in RMSE (both averaged across the original species and for O\textsubscript{3} individually) for each $\epsilon$ as compared to a value of zero, the fraction of simulations successfully completed (the ``stable solution rate''), and model calibration $R^2$ (Figure~\ref{fig:buffer_term}). 
As $\epsilon$ increases, the stable solution rate goes up, but the error also increases (Figure~\ref{fig:buffer_term}), demonstrating a tradeoff between the ability of the buffer term to prevent runaway error and its potential to interfere with the normal operation of the model.
The value of $\epsilon$ does not have a strong impact on model calibration on O\textsubscript{3} but decreases the average $R$-squared value of the latent species (Figure~\ref{fig:buffer_term}).

\subsection{Effect of Inclusion Probability Threshold}
\label{sec:IP} 
\begin{figure}[h]
\centering
 \centerline{\includegraphics[width=0.5\textwidth]{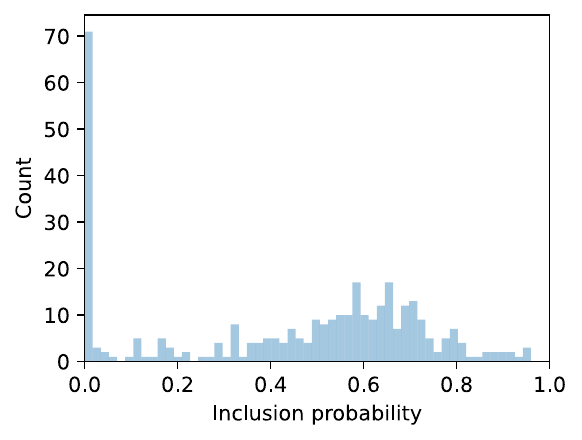}}
 \caption{Model coefficient inclusion probability distribution.}
 \label{fig:inclusionprobability}
 \end{figure}

As described in Section \ref{sec:esindy}, the inclusion probability for a given equation term from the candidate library ($\Theta$) is defined as the fraction of ensemble models that have a non-zero coefficient for that term (i.e. $\Xi_i \neq 0$ for term $i$).
Figure~\ref{fig:inclusionprobability} shows the distribution of inclusion probabilities for the $113 \times 3$ equation terms we consider in the selected 100~best-performing ensemble members. 
We find the inclusion probability has a bimodal distribution, with one mode at zero (for terms that are not included in any ensemble member) and a second mode centered at $~0.6$ (for terms that are included in some ensemble members but not others). 
Following work by \citeA{gao2023convergence}, we explore the possibility of removing some terms with low inclusion probability from the candidate library entirely.
We find that for $\lambda_{\mathrm{ip}}$ of 0, 0.5, and 0.7, the RMSE of the PPD mean for O\textsubscript{3} is 0.014~ppm, 0.032~ppm, and 0.104~ppm, respectively; and the stable solution rate is 66\%, 24\%, and 60\%, respectively.
Thus, we find that thresholding by inclusion probability is detrimental to model performance in this case and therefore we do not use it for the results described elsewhere in this analysis.
With different $\lambda_{\mathrm{ip}}$, we also explored the performance of the aggregated mean or median model, but found that these aggregate models are not be able to stably solve all the testing cases and do not improve accuracy as compared to the ensemble prediction or our selected ``best performing'' model.
 
\section{Discussion and Conclusion}

In this paper, we explore the use of the E-SINDy framework for creating a surrogate model of a simple gas-phase atmospheric chemical reaction system focused on tropospheric ozone formation, comprising 10 reactions, 11 species, and external inputs such as solar radiation, emission, and deposition. 
The E-SINDy surrogate model effectively facilitates uncertainty quantification and probabilistic predictions. This is exemplified by the PPD from E-SINDy, which accurately encompasses reference values throughout the trajectory for O\textsubscript{3}.

\citeA{xk2024surrogate} report that the development of a stable model necessitates substantial effort in finding a suitable candidate library. However, by utilizing a bootstrap sampling approach, we reduce the need for careful selection of the candidate library, thereby streamlining the process of surrogate model creation and producing a single ``best performing'' model which is more accurate than the one produced by deterministic SINDy, with additional accuracy improvement realized by the PPD mean prediction.

The ensemble model is able to produce well-calibrated PPDs for key species like H\textsubscript{2}O\textsubscript{2}, O\textsubscript{3}, and HO\textsubscript{2}, which closely align with reference values. 
Achieving a calibration $R$-squared value above 0.8 for uncertainty quantification results in a 3$\times$ slower inference with 30 ensemble members, but this approach promises greater efficiency when applied to more complex models.
Additionally, our experimentation with different sampling and postprocessing approaches led to the identification of an optimal, problem-specific bootstrapping method through hyperparameter tuning and the application of a coefficient inclusion probability threshold.

A main limitation of our study is that it only quantifies uncertainty in the simplification of a reference model—it does not consider uncertainty in the reference model with respect to reality.
This limitation could be overcome in future work by training surrogate models on a blend of data generated by reference models and observational data, for example collected in chamber experiments or sampling campaigns.
This would combine the volume and representativeness available from generated data with the accuracy of observational data.

Overall, results here pave the way for applying the E-SINDy framework to reference models characterized by higher stiffness, greater number and variability in species, and more complex reactions. 
Ultimately, we hope to scale the method here for operational use to simplify full-scale atmospheric chemical mechanisms for probabilistic prediction and uncertainty quantification in three-dimensional atmsopheric chemical transport models.
The variations observed in the distribution of concentration levels would then serve as an indicator of the uncertainty introduced by simplifying the detailed reference model into a surrogate model. 
Accordingly, this would only address uncertainty in the chemistry component relative to the reference model, but other model components could be surrogatized in the same way. 
The result would be an unprecedentedly comprehensive quantification of uncertainty within a model of atmospheric composition, leading to enhanced ability to understand the atmosphere and provide robust support for air quality control and regulation.

\section{Open Research}
All source code, including for dataset generation and for model training and evaluation is available through
Zenodo (\url{https://zenodo.org/records/12527214}).

\acknowledgments
The research is supported by NASA Early Career Faculty Award No. 80NSSC21K1813 and U.S. Environmental Protection Agency Grant No. R840012. 
It has not been formally reviewed by EPA. The views expressed in this document are solely those of the authors and do not necessarily reflect those of the Agency. 
The authors declare no conflicts of interest relevant to this study.

\bibliography{agusample.bib}
\end{document}


%
%


\title{Supporting Information for "Atmospheric Chemistry Surrogate Modeling with Uncertainty Quantification using Ensemble SINDy"}
%
%

%
%



\authors{
    \begin{minipage}{\linewidth}
Lin Guo\affil{1}, Xiaokai Yang\affil{1},  Zhonghua Zheng\affil{2}, Nicole Riemer\affil{3}, Christopher W. Tessum\affil{1}
    \end{minipage}
}


\affiliation{1}{Department of Civil and Environmental Engineering, University of Illinois Urbana-Champaign, Urbana, IL}
\affiliation{2}{Department of Earth and Environmental Sciences, The University of Manchester, Manchester, United Kingdom}
\affiliation{3}{Department of Climate, Meteorology \& Atmospheric Sciences, University of Illinois Urbana-Champaign, Urbana, IL}

%
%

%

\begin{article}

%
%

\noindent\textbf{Contents of this file}
\begin{enumerate}
\item Figures S1 to S11
\item Tables S1
\end{enumerate}

\noindent\textbf{Introduction}

1. Figure S1 to S11 are the Ensemble SINDy-based surrogate model performance for chemical
species prediction on the testing dataset;

2. Table S1 contains the reactions in the reference chemical mechanism.
\end{article}
\clearpage


 \begin{figure}
\centering
 \includegraphics[width=\textwidth]{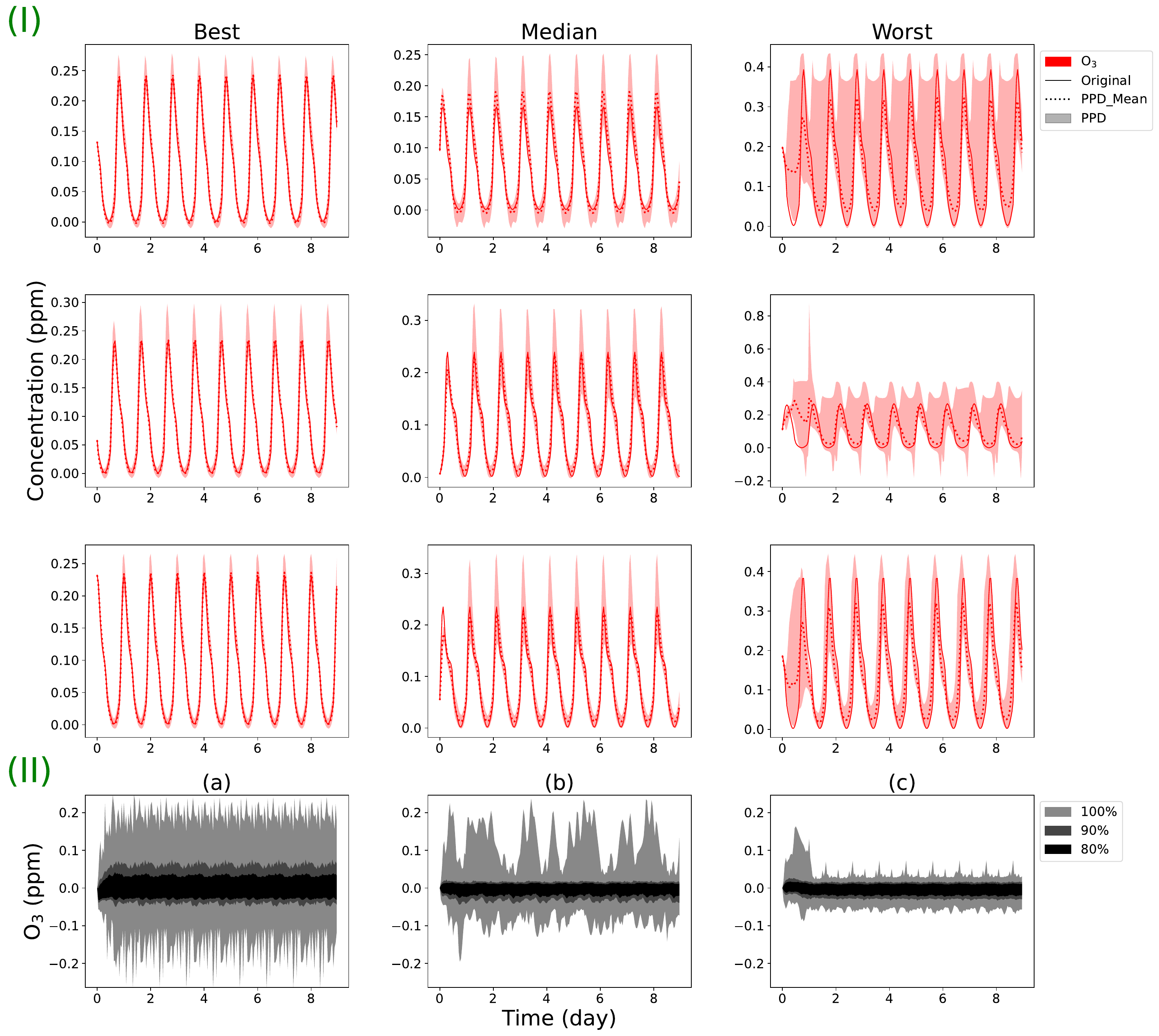}
 \caption{(I) The best, median and worst prediction on O\textsubscript{3} for nine-day testing cases (II) (a) Error percentile along nine days for O\textsubscript{3} among 375 testing cases by SINDy (b) Error percentile along nine days for O\textsubscript{3} among 375 testing cases by the best ensemble member (c) Error percentile along nine days for O\textsubscript{3} among 375 testing cases by the 100 selected ensemble members }
 \end{figure}

 \begin{figure}
\centering
 \includegraphics[width=\textwidth]{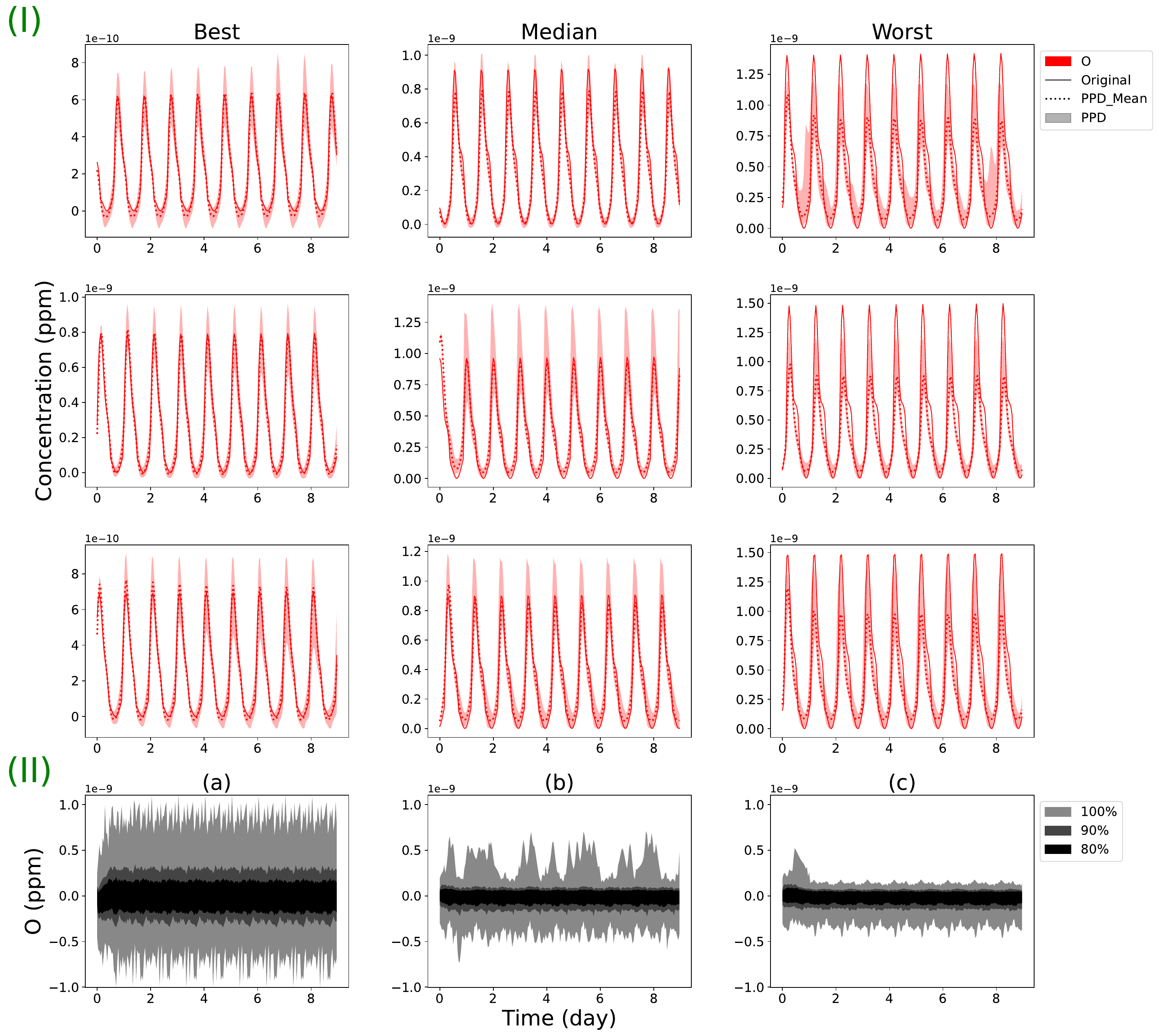}
 \caption{(I) The best, median and worst prediction on O for nine-day testing cases (II) (a) Error percentile along nine days for O among 375 testing cases by SINDy (b) Error percentile along nine days for O among 375 testing cases by the best ensemble member (c) Error percentile along nine days for O among 375 testing cases by the 100 selected ensemble members }
 \end{figure}

  \begin{figure}
\centering
 \includegraphics[width=\textwidth]{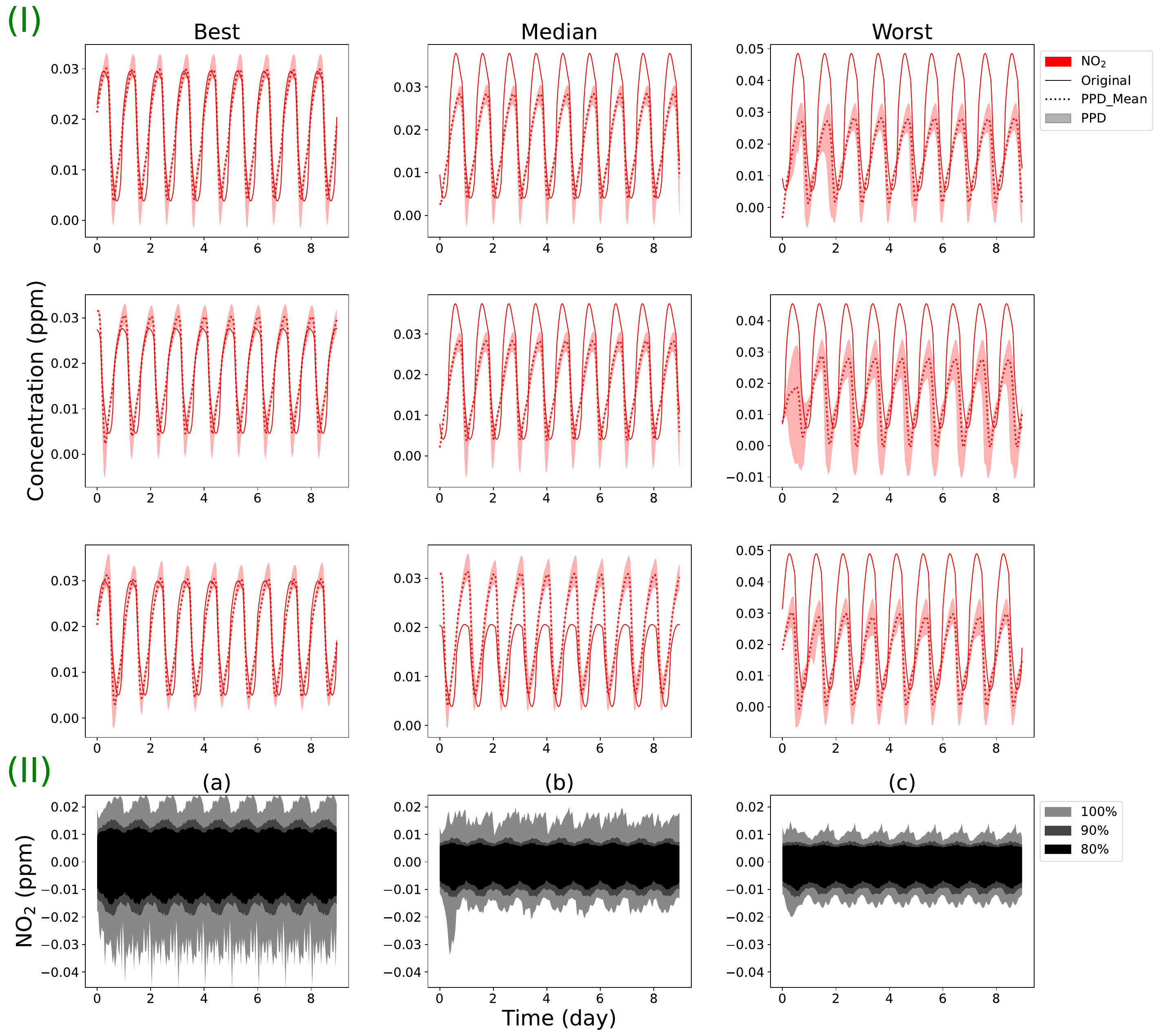}
 \caption{(I) The best, median and worst prediction on NO\textsubscript{2} for nine-day testing cases (II) (a) Error percentile along nine days for NO\textsubscript{2} among 375 testing cases by SINDy (b) Error percentile along nine days for NO\textsubscript{2} among 375 testing cases by the best ensemble member (c) Error percentile along nine days for NO\textsubscript{2} among 375 testing cases by the 100 selected ensemble members }
 \end{figure}

  \begin{figure}
\centering
 \includegraphics[width=\textwidth]{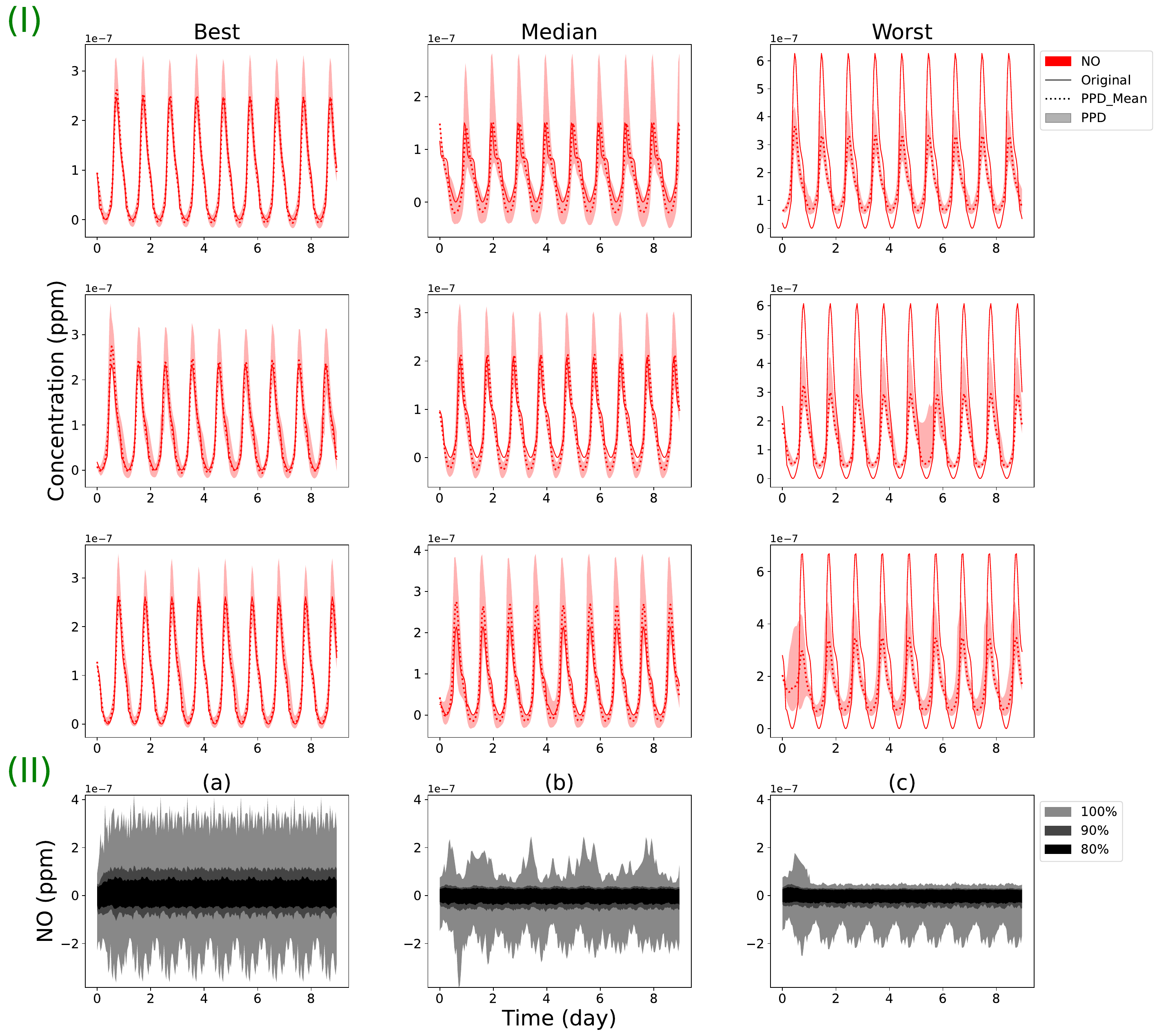}
 \caption{(I) The best, median and worst prediction on NO for nine-day testing cases (II) (a) Error percentile along nine days for NO among 375 testing cases by SINDy (b) Error percentile along nine days for NO among 375 testing cases by the best ensemble member (c) Error percentile along nine days for NO among 375 testing cases by the 100 selected ensemble members }
 \end{figure}

  \begin{figure}
\centering
 \includegraphics[width=\textwidth]{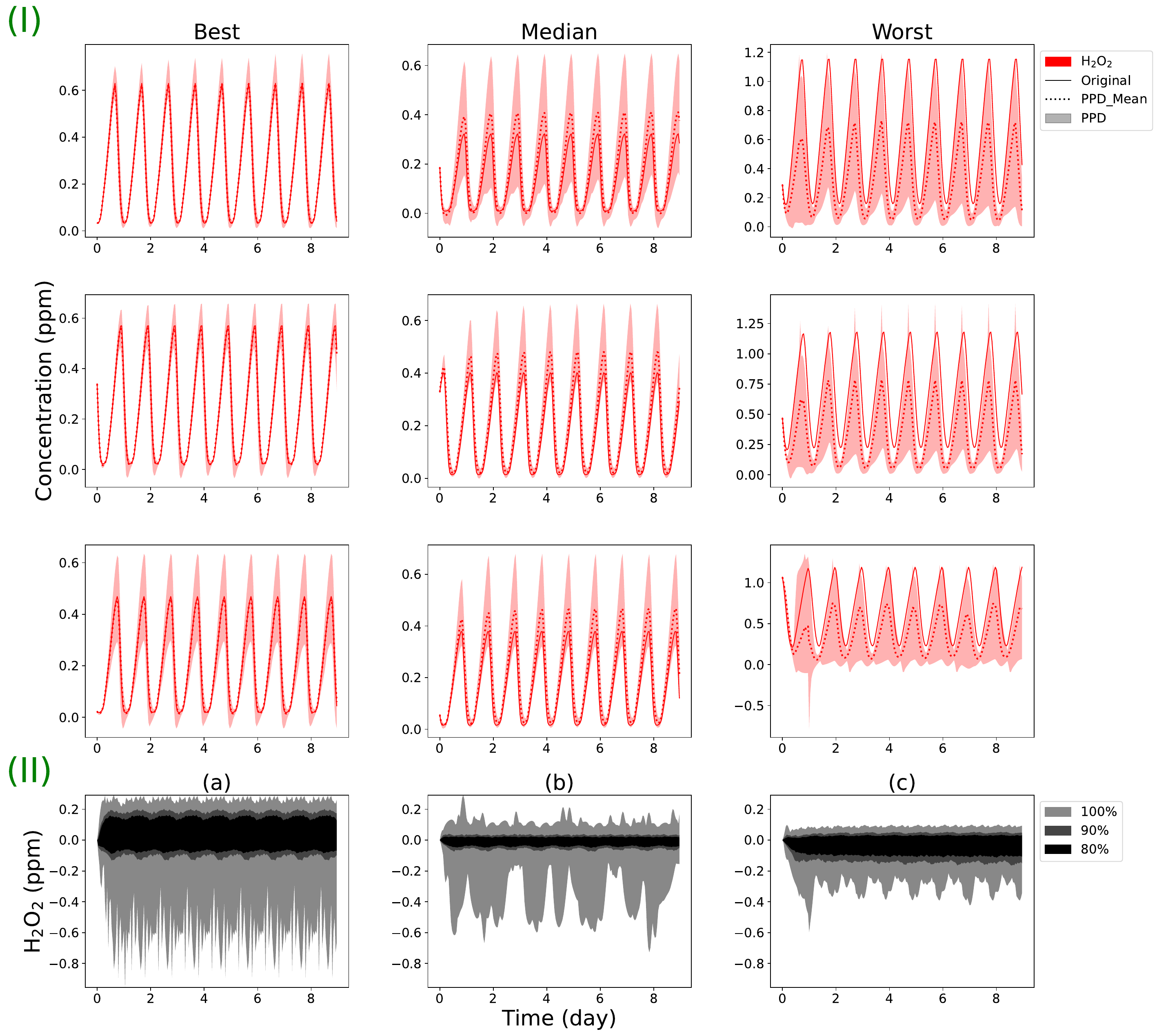}
 \caption{(I) The best, median and worst prediction on HO\textsubscript{2}H for nine-day testing cases (II) (a) Error percentile along nine days for HO\textsubscript{2}H among 375 testing cases by SINDy (b) Error percentile along nine days for HO\textsubscript{2}H among 375 testing cases by the best ensemble member (c) Error percentile along nine days for HO\textsubscript{2}H among 375 testing cases by the 100 selected ensemble members }
 \end{figure}

  \begin{figure}
\centering
 \includegraphics[width=\textwidth]{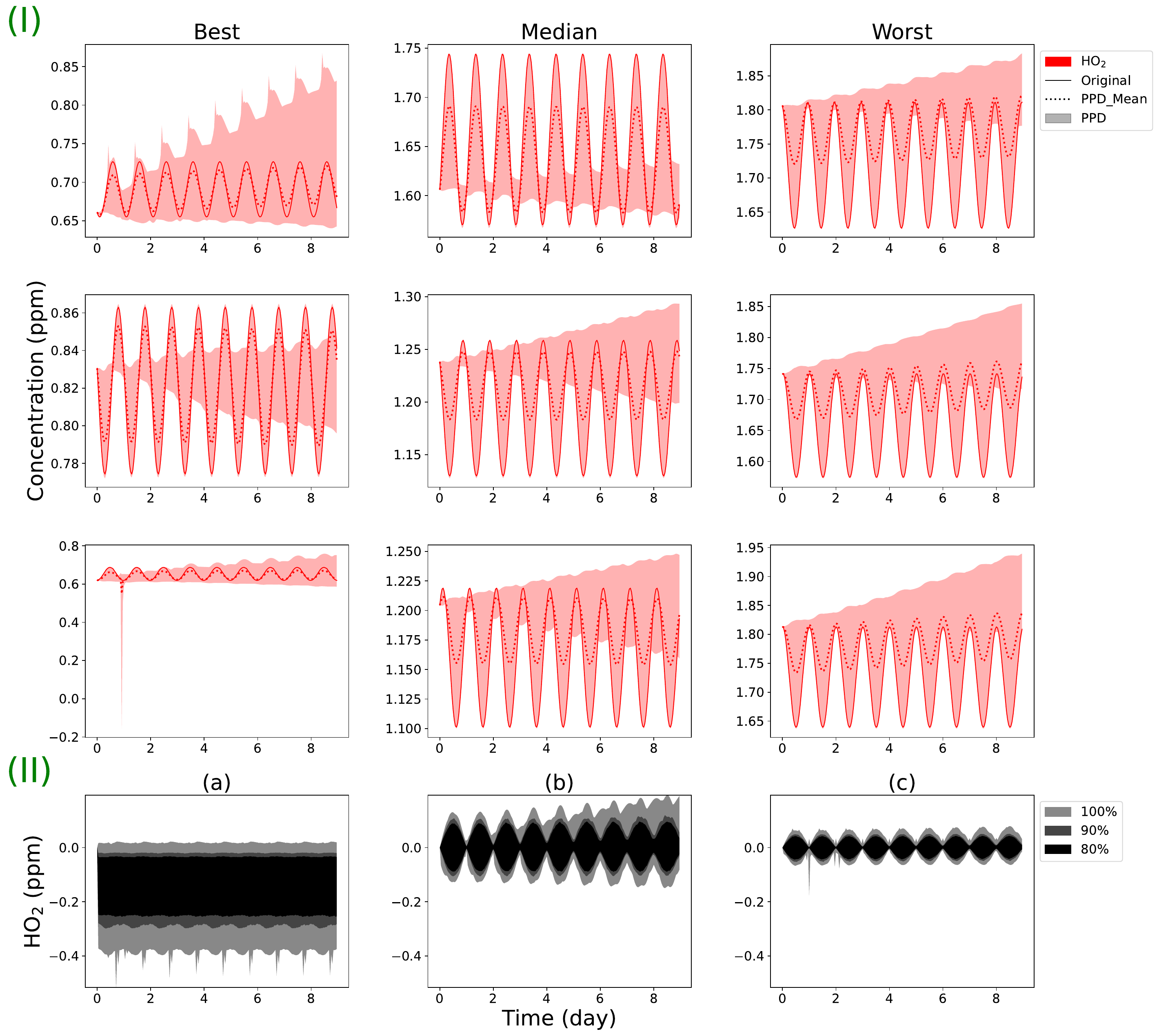}
 \caption{(I) The best, median and worst prediction on HO\textsubscript{2} for nine-day testing cases (II) (a) Error percentile along nine days for HO\textsubscript{2} among 375 testing cases by SINDy (b) Error percentile along nine days for HO\textsubscript{2} among 375 testing cases by the best ensemble member (c) Error percentile along nine days for HO\textsubscript{2} among 375 testing cases by the 100 selected ensemble members }
 \end{figure}

  \begin{figure}
\centering
 \includegraphics[width=\textwidth]{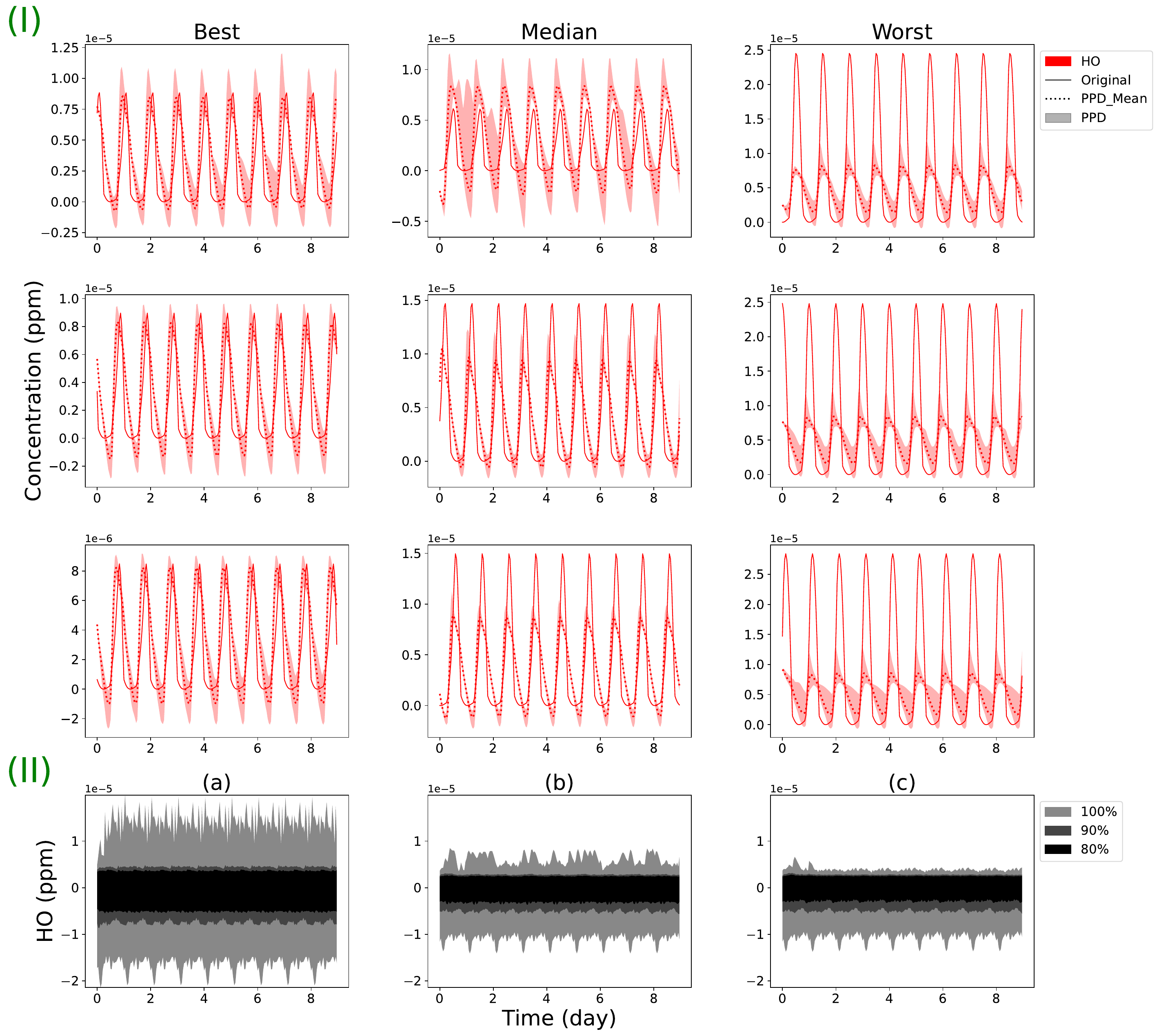}
 \caption{(I) The best, median and worst prediction on HO for nine-day testing cases (II) (a) Error percentile along nine days for HO among 375 testing cases by SINDy (b) Error percentile along nine days for HO among 375 testing cases by the best ensemble member (c) Error percentile along nine days for HO among 375 testing cases by the 100 selected ensemble members }
 \end{figure}

  \begin{figure}
\centering
 \includegraphics[width=\textwidth]{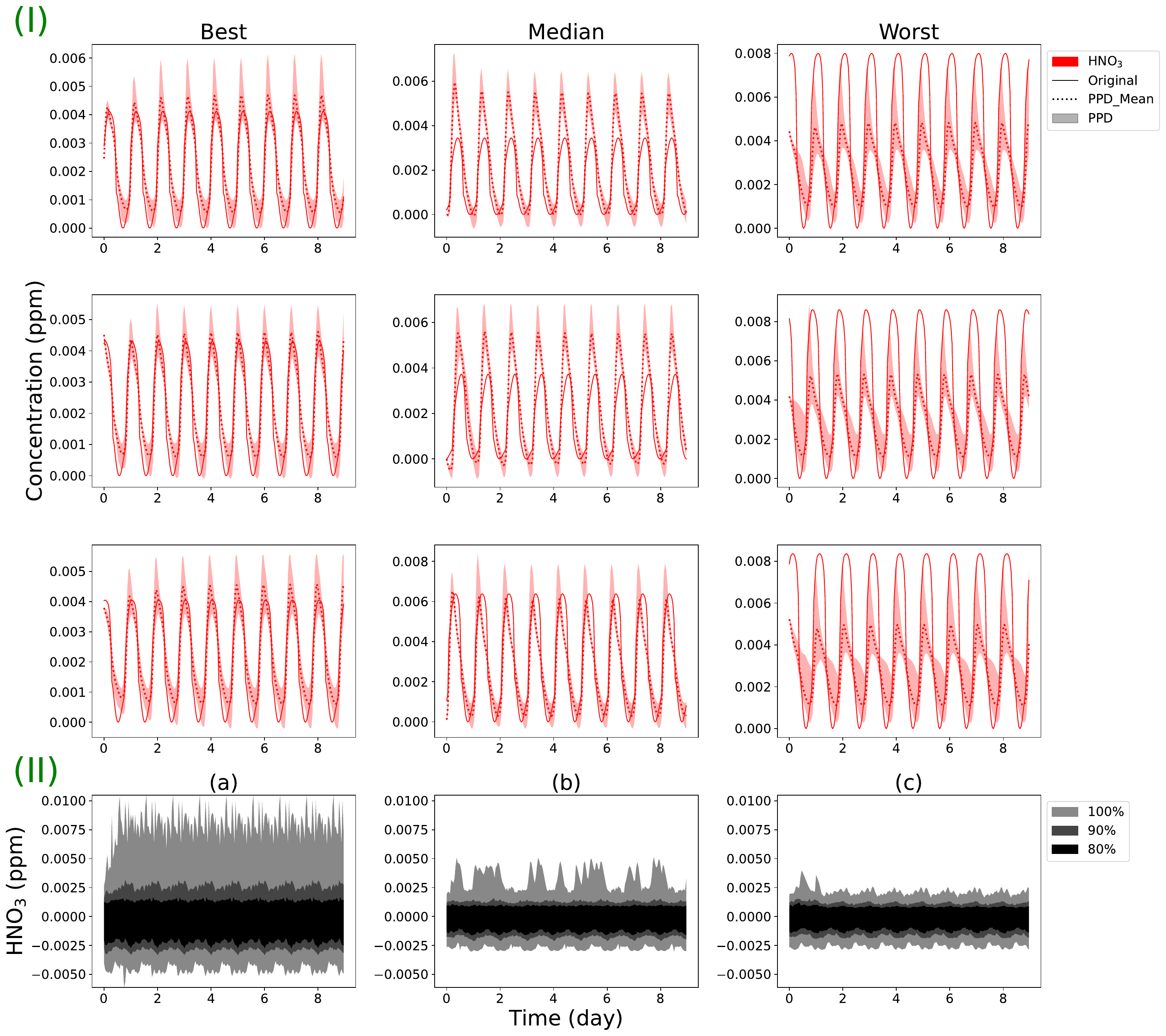}
 \caption{(I) The best, median and worst prediction on HNO\textsubscript{3} for nine-day testing cases (II) (a) Error percentile along nine days for HNO\textsubscript{3} among 375 testing cases by SINDy (b) Error percentile along nine days for HNO\textsubscript{3} among 375 testing cases by the best ensemble member (c) Error percentile along nine days for HNO\textsubscript{3} among 375 testing cases by the 100 selected ensemble members }
 \end{figure}

  \begin{figure}
\centering
 \includegraphics[width=\textwidth]{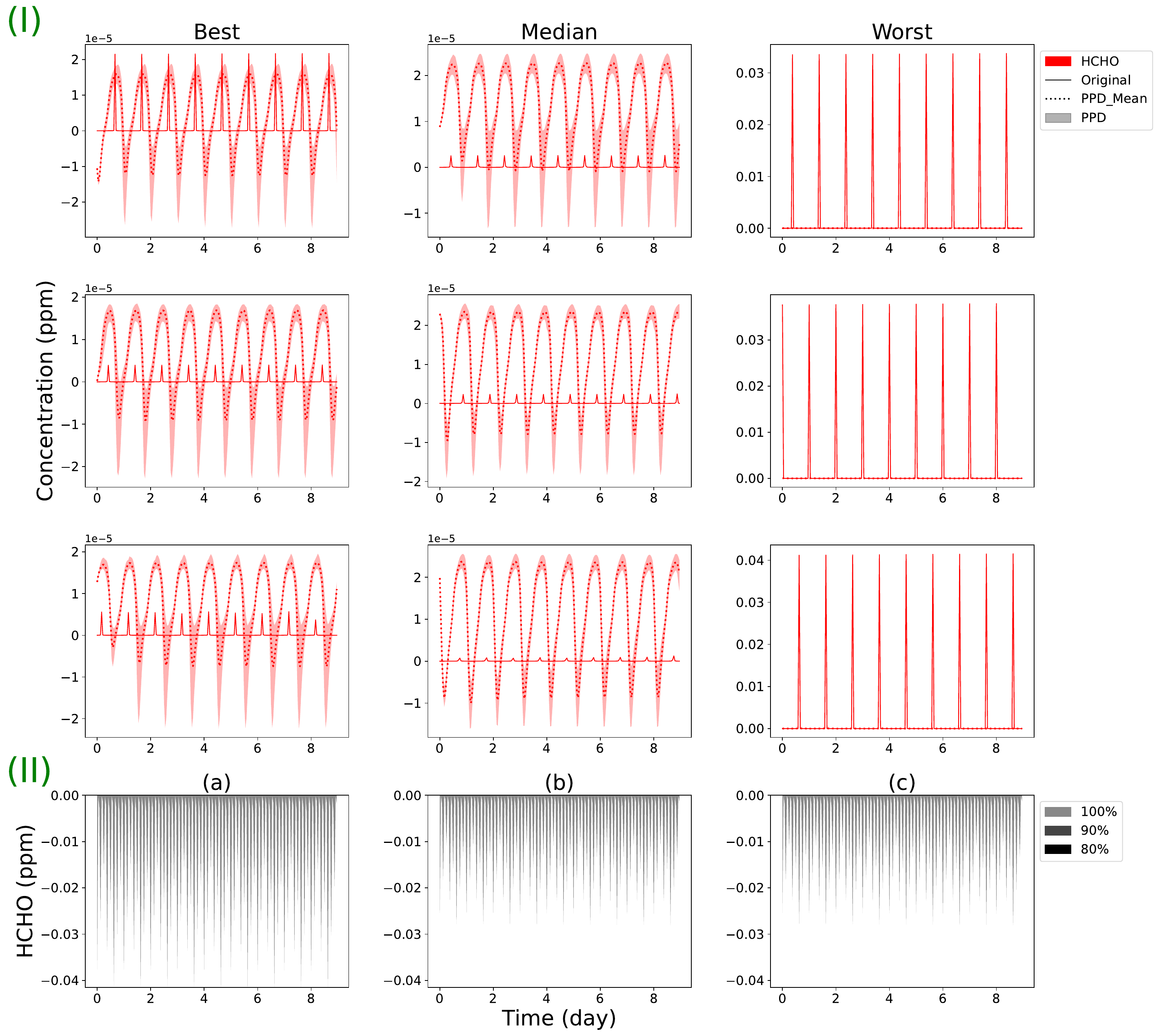}
 \caption{(I) The best, median and worst prediction on HCHO for nine-day testing cases (II) (a) Error percentile along nine days for HCHO among 375 testing cases by SINDy (b) Error percentile along nine days for HCHO among 375 testing cases by the best ensemble member (c) Error percentile along nine days for HCHO among 375 testing cases by the 100 selected ensemble members }
 \end{figure}

  \begin{figure}
\centering
 \includegraphics[width=\textwidth]{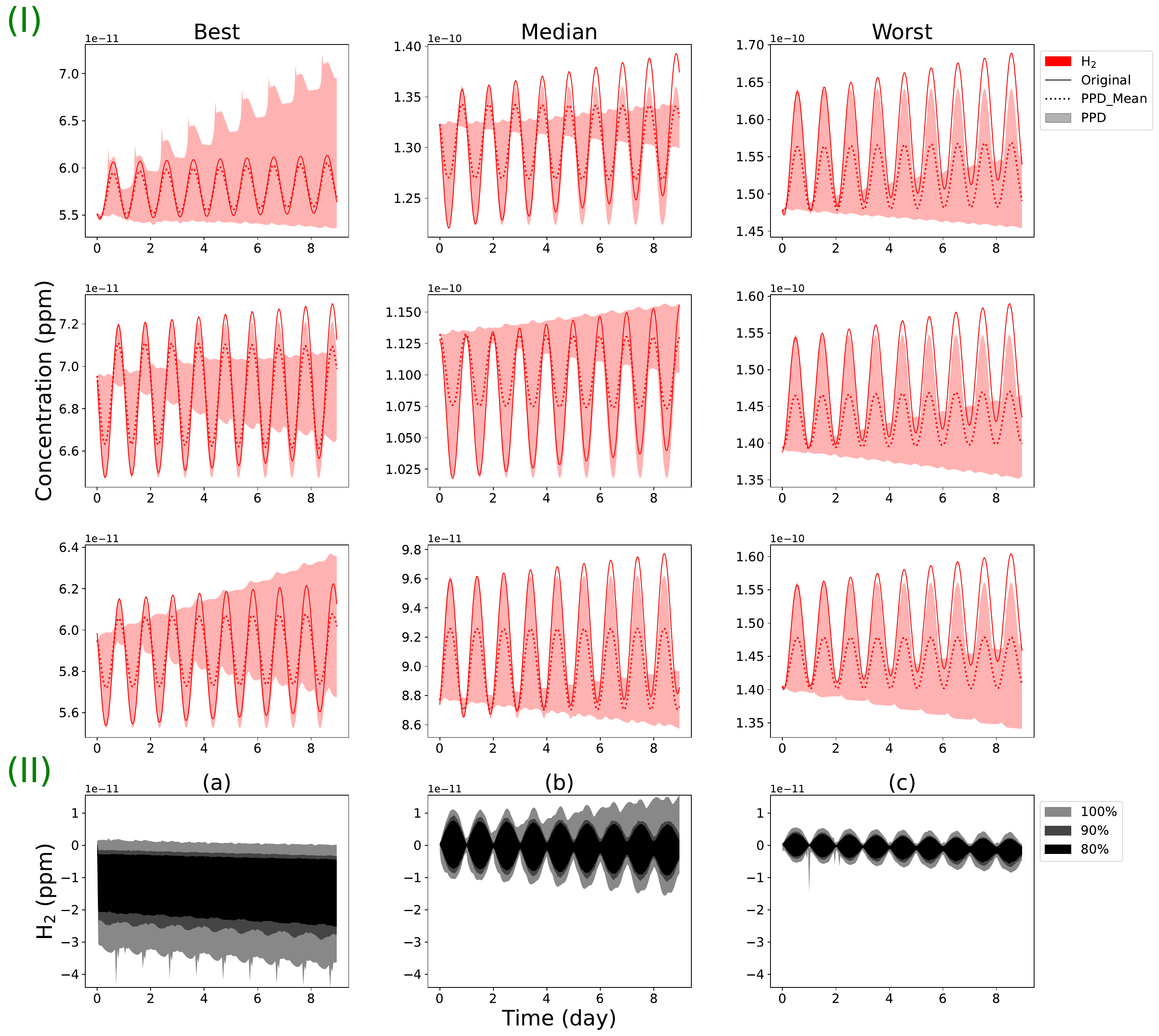}
 \caption{(I) The best, median and worst prediction on H\textsubscript{2} for nine-day testing cases (II) (a) Error percentile along nine days for H\textsubscript{2} among 375 testing cases by SINDy (b) Error percentile along nine days for H\textsubscript{2} among 375 testing cases by the best ensemble member (c) Error percentile along nine days for H\textsubscript{2} among 375 testing cases by the 100 selected ensemble members }
 \end{figure}

  \begin{figure}
\centering
 \includegraphics[width=\textwidth]{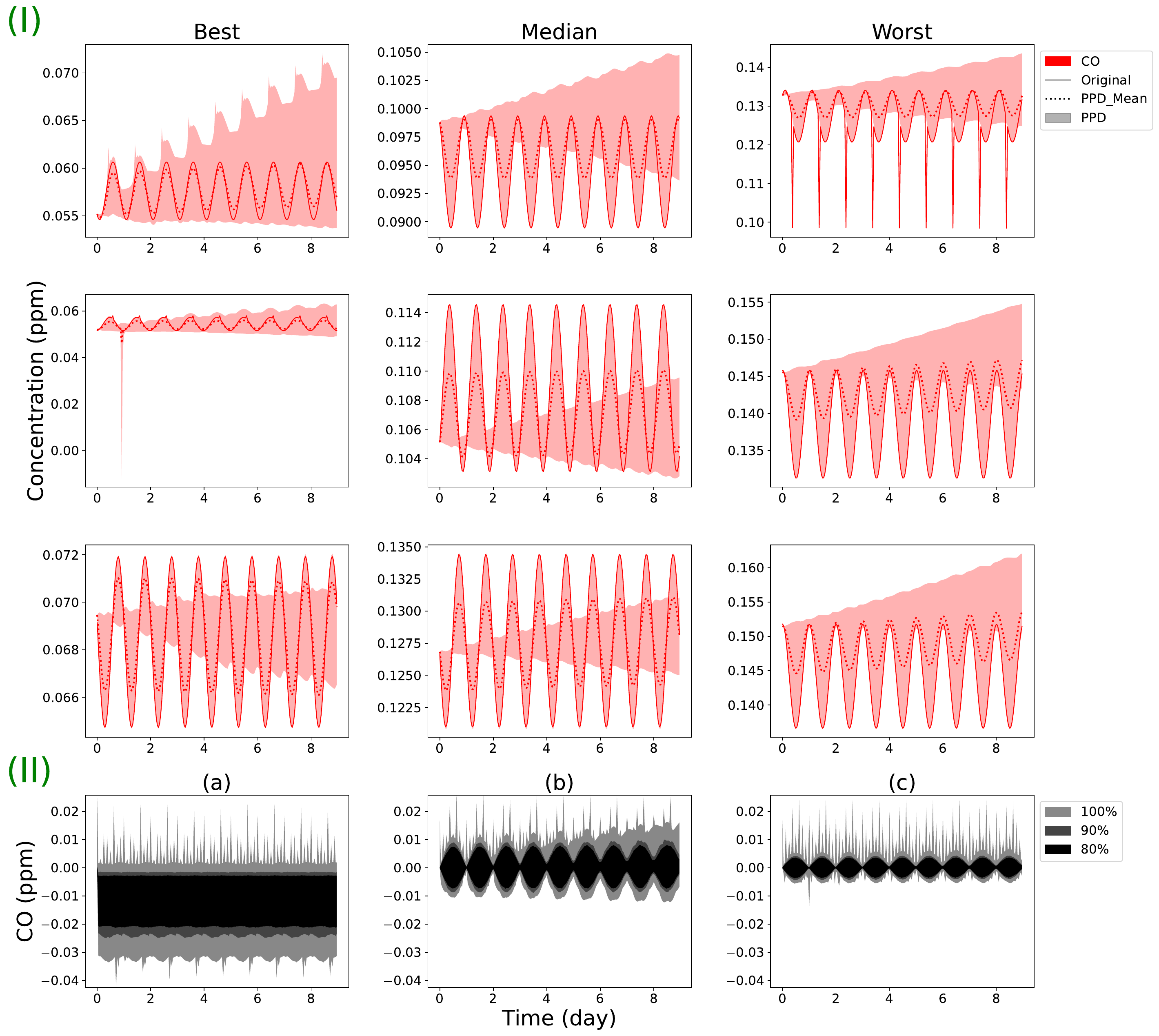}
 \caption{(I) The best, median and worst prediction on CO for nine-day testing cases (II) (a) Error percentile along nine days for CO among 375 testing cases by SINDy (b) Error percentile along nine days for CO among 375 testing cases by the best ensemble member (c) Error percentile along nine days for CO among 375 testing cases by the 100 selected ensemble members }
 \end{figure}

\begin{table}
\settablenum{S1} 
\caption{Reactions in reference chemical mechanism}
\centering
\begin{tabular}{l c}
\hline
Number& Reaction                         \\
\hline
(1)    & NO\textsubscript{2} + hv $\rightarrow$ NO + O\\
(2)    & O + O\textsubscript{2}  $\rightarrow$ O\textsubscript{3}\\
(3)    & O\textsubscript{3} + NO  $\rightarrow$ NO\textsubscript{2} + O\textsubscript{2}\\
(4)    & HCHO + hv  $\rightarrow$ 2 HO\textsubscript{2} + CO\\
(5)    & HCHO + hv  $\rightarrow$ H\textsubscript{2} + CO\\
(6)    & HCHO + OH  $\rightarrow$ HO\textsubscript{2} + CO + H\textsubscript{2}O\\
(7)    & HO\textsubscript{2} + NO  $\rightarrow$ OH + NO\textsubscript{2}\\
(8)    & OH + NO\textsubscript{2}  $\rightarrow$ HNO\textsubscript{3}\\
(9)    & H\textsubscript{2}O\textsubscript{2} + hv  $\rightarrow$ 2 OH\\
(10)   & H\textsubscript{2}O\textsubscript{2} + OH  $\rightarrow$ H\textsubscript{2}O + HO\textsubscript{2}\\
\hline
\end{tabular}
\end{table}